\RequirePackage{fix-cm}
\documentclass[smallextended]{svjour3}       
\smartqed  
\usepackage{graphicx}
\usepackage{amsmath,amssymb,latexsym}
\usepackage{hyperref}
\journalname{General Relativity and Gravitation}
\begin{document}

\title{On Shearing Fluids with Homogeneous Densities\\
  \thanks{The final publication is available at Springer via http://dx.doi.org/[10.1007/s10714-016-2065-y].}
}



\author{D.C.Srivastava \and 
                V.C.Srivastava \and Rajesh Kumar 
}


\institute{D.C.Srivastava \at
            Retd,  Department of Physics, DDU Gorakhpur University, Gorakhpur - 273009, U. P. India\\
              \email{dcs.gkp@gmail.com}           
           \and
           V.C.Srivastava \at
                Department of Physics, DDU Gorakhpur University, Gorakhpur- 273009, U. P. India\\
                 \email{4marchvcs@gmail.com} 
	\and
	Rajesh Kumar \at
	Department of Mathematics and Statistics, DDU Gorakhpur University, Gorakhpur- 273009, U. P. India\\
}

\date{Received: date / Accepted: date}

\maketitle

\begin{abstract}
In this paper, we study shearing spherically symmetric homogeneous density fluids in comoving coordinates. It is found that the expansion of the four-velocity of a perfect fluid is homogeneous, whereas its shear is generated by an arbitrary function of time $M(t)$, related to the mass function of the distribution. This function is found to bear a functional relationship with density. The field equations are reduced to two coupled first order ordinary differential equations for the metric coefficients $g_{11}$ and $g_{22}$. We have explored a class of solutions assuming that $M$ is a linear function of the density. This class embodies, as a subcase, the complete class of shear-free solutions. We have discussed the off quoted work of Kustaanheimo (1947) and have noted that it deals with shear-free fluids having anisotropic pressure. It is shown that the anisotropy of the fluid is characterized by an arbitrary function of time. We have discussed some issues of historical priorities and credentials related to shear-free solutions. Recent controversial claims by Mitra (2011, 2012) have also been addressed. We found that the singularity and the shearing motion of the fluid are closely related. Hence, there is a need for fresh look to the solutions obtained earlier in comoving coordinates.
\end{abstract}

\section{Introduction}\label{sec1}
Oppenheimer and Snyder \cite{os39} in a pioneering work in 1939 considered the problem of the gravitational collapse of a star in the last stages of its evolution, when pressure equilibrium is no longer possible because of the exhaustion of the star's thermonuclear fuel. They discussed the problem in a simplified case of a non-interacting perfect fluid. Since then, possible generalizations to more realistic situations have been examined. Impetus was given to research in this area by the discovery of quasars in the early 1960s. To quote Misner and Sharp \cite{ms64}, the objective of such research is to explore \textit{`` the possibility (in a stage of collapse where the gravitational binding energy $GM^{2}/ R$  becomes comparable to the rest energy $M c^{2}$) of a large energy output of a star.}"  Black hole formation, singularity theorems and related issues are topics of current interest in the area.\par
Spherically symmetric perfect fluid distributions having homogeneous densities have been studied in the literature under some simplifying assumptions viz., (i) isotropic coordinates, (ii) comoving coordinates, (iii) shear-free motion (iv) regularity of the solution at the origin of the coordinate system. Either of the assumptions of  isotropic coordinates and of comoving coordinates can always be introduced with out any loss of generality. But  both employed together restrict the motion  to be shear-free, Kramer $et~al$.\cite{skmh80}. Similarly the requirement of regularity of the solution at the origin has been found to restrict the motion to be shear-free (Thompson and Whithrow \cite{tw67} and Misra and Srivastava \cite{rd72}, \cite{rd73}).\par
The shear-free assumption has been utilized often in the studies of radiating stars and of gravitational collapse. This assumption coupled with the condition of homogeneous expansion rate is equivalent to the homology conditions in the Newtonian limit. Studies of radiating star models with dissipating processes, and with anisotropic fluids are among the current problems in the field. Heat flux is considered important for various astrophysical problems viz, modelling in structure formation, evolution of voids, gravitational collapse, inhomogeneous cosmologies, singularities, and black hole physics, Krasinski \cite{ka06}. Herrera $et~al.$\cite{hpo06} have shown that heat flow is a requirement in thermal evolution of the collapsing sphere modelled in causal thermodynamics. Heat conducting and collapsing shear-free fluids have been considered recently by Ivanov \cite{i12}, Msomi $et~al.$\cite{mgm11}, and Nyonyi $et~al.$(\cite{nmg13}, \cite{nmg14}). The stability of shear-free collapse has also been examined. Herrera $et~al.$ \cite{hs10} have demonstrated that pressure anisotropy and dissipation affect the shear-free condition, and an initially shear-free gravitating relativistic fluid can become unstable. It is worth to point out that in these studies the Lie symmetry approach has been applied successfully for obtaining the solutions of boundary junction condition equations, and also for generating models, notably the expanding, accelerating and shearing models (Msomi $et~al.$\cite{mgm11}, Abebe $et~al.$\cite{amg14}, \cite{amg15}).\par
 McVittie and Wiltshire \cite{mw77} put forward an idea in 1977 to look for  solutions in noncomoving coordinates with the argument that  a simple looking solution in one coordinate system may look bewildering when transformed to some other coordinate systems. By that time, it was realized that on one hand there is a need to discuss more general motion, in particular shearing motion,  whereas on other hand almost all the solutions obtained in comoving coordinate systems represent shear-free motion of the perfect fluid, Kramer $et~al.$ \cite{skmh80}. Hence, studies of solutions representing shearing solutions and / or obtained using noncomoving coordinates  soon picked up the momentum, to name a few, Ray \cite{r78} , Van Den Bergh and Wils \cite{bw85}, Maharaj $et~al.$\cite{mmm93}, Bonnor and Knutsen \cite{bk93}, Knutsen \cite{kh95}, Davidson \cite{d03}, Wiltshire \cite{w06}. The solutions discovered in noncomoving coordinates are mostly been found to have the singularity at $r=0$ and are considered as to serve models for early / and late stages of Gravitational collapse and/or Cosmological type singularity. This has motivated us to take up a project of studying fluid distributions in comoving coordinates. In the present investigation we are concerned with homogeneous density distribution. This topic as such is very old and has been studied extensively(\cite{tw67} - \cite{rd73}, \cite{kp47} - \cite{tah68}). We have preferred to base the investigation on the one carried out by one of us long back (Misra and Srivastava \cite{rd72}, \cite{rd73}). However, they focused on perfect fluid spheres and their objective was to establish that the regularity conditions related to elementary flatness at the centre require shear of the four-velocity to vanish.
\par                                                                    
Recently, Durgapal and Fuloria \cite{df10}  have done an analysis considering baryonic conservation law and no heat transfer condition. They found that within the assumed conditions of shear-free motion and the boundary conditions the only valid solution for the collapse of a uniform density sphere is the solution given by Oppenheimer and Snyder \cite{os39}. Mitra \cite{ma12}  also considered uniform density and shear-free motion of perfect fluid and has exclaimed :\textit{~~``...  in the past 65 years of innumerable authors working on this problem failed to see that the collapse of a supposed homogeneous sphere is (actually)  synonymous to the old Oppenheimer- Snyder problem}". Besides, he places his result  supporting to the similar finding by Durgapal and Fuloria \cite{df10}. In another marginally earlier paper, Mitra \cite{ma11} had also claimed to have established this finding.
\par
Basic equations are given in the next Section. In Sect.\ref{sec3} we have shown that  the class of shearing solutions are determined subject to the integration of two first order coupled ordinary differential equations for two of the metric coefficients. This class of solutions contains four arbitrary functions of integrations; $F(t)$, $G(t)$, $Q(t)$ and $M(t)$. The function $F(t)$ represents the expansion of the four-velocity of the distribution, whereas the function $G(t)$ characterizes the shear of the four-velocity vector. The function $Q(t)$ is related linearly with $\rho (t)$ and $F^{2}(t)$. The function $M(t)$ characterizes the mass function $m(r, t)$ of the distribution. It is found that the functions $M(t)$ and $\rho (t)$  bear a functional relationship. $M(\rho)$ is interpreted as the generator of the shear of the four- velocity of the fluid; for shear-free motion $M$ is a constant.
\par
In Sect.\ref{sec4} we present a class of shearing solutions obtained under the assumption that there is a linear relationship between $M$ and $\rho$. It is shown that its particular case $M$=constant represents the class of all shear-free solutions. Krasinski \cite{ka06} has identified the class of shear-free solutions as Qvist / Kustaanheimo class of solutions, Qvist \cite{qb47}, Kustaanheimo \cite{kp47}. We have found unjustified to give credit to Qvist \cite{qb47}. Kustaanheimo \cite{kp47} first considered the fluid with anisotropic pressure and at a later stage introduced the assumption of isotropy of pressure. Accordingly, we analyse in Sect.\ref{sec5} shear-free fluids with  anisotropic pressure. The anisotropy of the fluid distribution is found to be characterized by an arbitrary function of time. In Sect.\ref{sec6} we present a discussion of the function $M(t)$ along with our concluding remarks. In this Section we also include a discussion of some issues of historical credentials related to the shear-free class of solutions, and also about Mitra's analyses (\cite{ma12},\cite{ma11}).\\

\section{Basic Equations}\label{sec2}
The spherically symmetric fluid distribution is described by the  metric
\begin{equation}
ds^{2}=  e^{\nu}~dt^{2} -  e^{\lambda}dr^{2} - e^{\mu}[d\theta^{2} + (\sin^{2}\theta)~d\phi^{2}]
\label{eq1}
\end{equation}
where $\lambda$, $\mu$ and $\nu$ are functions of ($r$, $t$) only. We assume the coordinate system $x^{i}$:($t$, $r$, $\theta$, $\phi$) to be comoving with the matter. The four-velocity vector  $V^{i}$ and the related kinematical quantities are given as
\begin{equation}
V^{i} = (e^{\frac{-\nu}{2}}, 0, 0, 0),~~~~ a^{i}=(0, \frac{\nu'}{2}, 0, 0)
\label{eq2}
\end{equation}
\begin{equation}
 \omega = 0,\quad  \Theta = e^{\frac{-\nu}{2}}(\frac{\dot{\lambda}}{2}+\dot{\mu}),\quad {\sigma}= |\frac{1}{2\sqrt{3}} e^{-{\nu}/2} ({\dot{\lambda}}-{\dot{\mu}})|
\label{eq3001}
\end{equation}
The nonvanishing components of the  energy momentum tensor $T^{i}\ _{j}$ for the fluid distribution are given as
\begin{equation}
T^{0}\ _{0}=\rho(r, t) ;~\quad T^{1}\ _{1}=-p_{1}(r, t),~~ T^{2}\ _{2}=T^{3}\ _{3}=-p_{2}(r, t)
\label{eq3}
\end{equation}
where $\rho$ represents the matter density, and $p_{1}$ and $p_{2}$ represent the pressures of the fluid. Einstein's field equations may now be set up as
\begin{equation}
-e^{-\lambda}(\frac{\mu'^{2}}{4}+\frac{\mu'\nu'}{2})+e^{-\mu}+e^{-\nu}(\ddot{\mu}-\frac{\dot{\mu}{\dot{\nu}}}{2}+\frac{3\dot{\mu^{2}}}{4})=-8\pi p_{1}
\label{eq4}
\end{equation}
\begin{multline}
-\frac{e^{-\lambda}}{4} (2\mu''+2\nu''+\mu'^{2}+\nu'^{2}+\mu'\nu'-\lambda'\nu'-\lambda'\mu')\\\
+\frac{e^{-\nu}}{4}(2\ddot{\mu}+2\ddot{\lambda}+\dot{\mu}^{2}+\dot{\lambda}^{2}+\dot{\mu}\dot{\lambda}-\dot{\mu}\dot{\nu}-\dot{\lambda}\dot{\nu})=-8\pi p_{2}
\label{eq5}
\end{multline}

\begin{equation}
-e^{-\lambda}(\mu''+\frac{3\mu'^{2}}{4}-\frac{\lambda'\mu'}{2})+e^{-\mu}+e^{-\nu}(\frac{\dot{\mu}^{2}}{4}+\frac{\dot{\lambda}\dot{\mu}}{2})=8\pi\rho
\label{eq6}
\end{equation}
\begin{equation}
2\dot{\mu}'+\dot{\mu}\mu'-\dot{\lambda}\mu'-\dot{\mu}\nu'=0
\label{eq7}
\end{equation}\\
Here and onwards, an overhead dot and a prime denote the partial derivatives with respect to $t$ and $r$, respectively.\par
We, hereafter, unless state otherwise, specialize the analysis to the case of a perfect fluid i.e. assume
\begin{equation}
p_{1}=p_{2}=p~(Say)
\end{equation}
 Equations (\ref{eq4})-(\ref{eq7}),  because of the Bianchi Identities, are not independent of each other. Accordingly, one obtains
\begin{equation}
p'=-\frac{\nu'(p+\rho)}{2}
\label{eq9}
\end{equation}
\begin{equation}
\dot\rho=-{(\dot\mu+\frac{\dot\lambda}{2})(p+\rho)}
\label{eq10}
\end{equation}
The left hand sides of Eqs (\ref{eq4}), (\ref{eq6}) and (\ref{eq7}) have a very elegant  and useful symmetrical structure. Let us introduce, following Misner and Sharp \cite{ms64},  a variable $m(r,t)$ defined as
\begin{equation}
8m(r,t)= \dot{\mu}^{2}e^{\frac{3\mu}{2}-\nu}+ 4e^{\frac{\mu}{2}}-\mu'^{2}e^{\frac{3\mu}{2}-\lambda}
\label{eq11}
\end{equation}
This function has been recognized as the mass function of the distribution (Misner and Sharp \cite{ms64}, Thompson and Whithrow \cite{tw67}).  Equations (\ref{eq4}), and (\ref{eq6}) after making use of Eqs. (\ref{eq7}) and (\ref{eq11}) may be recast respectively as
\begin{equation}
\dot{m} = -2\pi p\dot{\mu} e^{3\mu/2}
\label{eq12}
\end{equation}

\begin{equation}
m' = 2\pi\rho\mu' e^{3\mu/2}
\label{eq13}
\end{equation}

The set of Eqs. (\ref{eq9}) - (\ref{eq13}) in the case $\dot{\mu} \neq 0$  is completely equivalent to the set of Eqs, (\ref{eq4}) - (\ref{eq7}). However, occasionally use of both set of equations turns out to be more convenient. It  must be noted that Eqs. (\ref{eq9}) and (\ref{eq10}) contain derivatives of the physical variables $p$  and $\rho$, and hence, their solutions need to be fed back into their respective parent equations, (\ref{eq4}) and (\ref{eq6}). Further, it is pointed out that the integrability condition of the function $m(r,t)$, after one employs Eqs. (\ref{eq9}) and (\ref{eq10}) to eliminate the derivatives of $p$ and $\rho$, leads to Eq.(\ref{eq7}).\par

\section{Class of Homogeneous Density Solutions}\label{sec3}
We assume that the density of the perfect fluid distribution does not vary across the time sections of the comoving system of the coordinates i.e.
\begin{equation}
\rho=\rho~(t)
\label{eq14}
\end{equation}
It is simple to integrate Eq.(\ref{eq9}) with respect to $‘r’$  and to get
\begin{equation}
e^{\frac{\nu}{2}}(p+\rho)= L(t)
\label{eq15}
\end{equation} 
where $L(t)$ is a constant of integration. Here, it is remarked that our analysis is to  be taken with the proviso that $\nu' \neq 0$. The case $\nu'$=0 requires a separate analysis  and has been dealt with in the literature in its full generality  (see for example,   Herlt \cite{he96} ).  Eq. (\ref{eq10}), after eliminating ($p$+ $\rho$) by using Eq. (\ref{eq15}), may be written as
\begin{equation}
e^{-\frac{\nu}{2}}(\dot{\mu}+\frac{\dot{\lambda}}{2})= -\frac{\dot{\rho}}{L(t)}\equiv F(t)
\label{eq16}
\end{equation} 
This in view of Eq.(\ref{eq3001}) reveals that the expansion $\Theta$ of the four-velocity vector  is uniform,
\begin{equation}
\Theta = F(t)
\label{eq17}
\end{equation}
Hence $F(t) \neq 0$,  otherwise  it will lead to a static configuration, a case not of our interest.  Equation (\ref{eq7}), after $\dot{\lambda}$ is eliminated using Eq.(\ref{eq16}), becomes integrable  with respect to $‘r’$, and one
obtains
\begin{equation}
e^{-\frac{\nu}{2}}\dot{\mu}= \frac{2F(t)}{3}+G(t)e^{\frac{-3\mu}{2}}
\label{eq18}
\end{equation} 
where $G(t)$ is an arbitrary constant of integration. Equations (\ref{eq16}) and (\ref{eq18}) lead to
\begin{equation}
e^{-\frac{\nu}{2}}\frac{\dot{\lambda}}{2} = \frac{F(t)}{3}-G(t)e^{\frac{-3\mu}{2}}
\label{eq19}
\end{equation} 

\begin{equation}
\dot{\lambda}=2\dot{\mu}\frac{[\frac{1}{3}-\mathcal{G}~e^{\frac{-3\mu}{2}}]}{[\frac{2}{3}+\mathcal{G}~e^{\frac{-3\mu}{2}}]},~\quad {\mathcal{G}(t) \equiv \frac{G(t)}{F(t)}}
\label{eq20}
\end{equation} 
Let us note that
\begin{equation}
G=0 \Leftrightarrow \dot{\lambda}=\dot{\mu}
\label{eq22}
\end{equation}\par
Equation (\ref{eq13}), in view of  Eq. (\ref{eq14}), is easily integrated with respect to $‘r’$ as 
\begin{equation}
m=\frac{4\pi}{3}\rho e^{\frac{3\mu}{2}} + M(t)
\label{eq23}
\end{equation} 
where $M(t)$ is a constant of integration.  Equations (\ref{eq12}),  (\ref{eq23}) and (\ref{eq10}) now yield
\begin{equation}
\frac{2\pi}{3} (1-\frac{\dot{\lambda}}{\dot{\mu}})~\dot{\rho}~e^{3\mu/2} = \dot{M(t)} (1+\frac{\dot{\lambda}}{2\dot{\mu}})
\label{eq24}
\end{equation}
This equation in view of Eq. (\ref{eq20}) leads to the result 
\begin{equation}
2\pi~{G(t)}~\dot{\rho} = \dot{M}~{F(t)}
\label{eq25}
\end{equation}
,indicating that there is a functional relationship between $M$ and $\rho$:
\begin{equation}
M=M(\rho),~~~~~ \frac{dM}{d\rho}=2\pi~\mathcal{G}(t)
\label{eq500}
\end{equation}
This equation means that
\begin{equation}
G(t)=0 \Leftrightarrow \mathcal{G}=0 \Leftrightarrow M(t) = Constant = M_{0}~~~~~(Say)
\end{equation}
Besides, we have to see the consequence of the result (\ref{eq23}) from another angle as well. Consider Eq.(\ref{eq23}) and substitute for (i)  $m(r, t)$ using its defining equation, (\ref{eq11}), and (ii) $\rho$ from Eq. (\ref{eq6}). Thereafter, eliminate $\dot{\mu}$ and $\dot{\lambda}$ from the resulting equation using Eqs. (\ref{eq18}) and (\ref{eq19}) to obtain
\begin{equation}
\frac{e^{\frac{3\mu}{2}-\lambda}}{12}(2\mu''-\lambda'\mu')+\frac{e^{\frac{\mu}{2}}}{3}+\frac{G^{2}(t)}{4}e^{\frac{-3\mu}{2}}= M(t)-\frac{1}{6}F(t)G(t)
\label{eq26}
\end{equation} 
This equation is easily integrated with respect to $‘r’$ to obtain
\begin{equation}
\frac{e^{-\lambda}\mu'^{2}}{12}-\frac{e^{-\mu}}{3}-\frac{G^{2}(t)e^{-3\mu}}{12}=-\frac{2}{3}[M(t)-\frac{G(t)F(t)}{6}]e^{\frac{-3\mu}{2}}+Q(t)
\label{eq27}
\end{equation} 
where $Q(t)$ is an arbitrary function of integration. Equation (\ref{eq6}) in view of Eqs. (\ref{eq18}), (\ref{eq19}), (\ref{eq26}) and (\ref{eq27}) yields the expression for the density as
\begin{equation}
8\pi\rho= \frac{F^{2}(t)}{3}-9~Q(t)
\label{eq28}
\end{equation}
This means $Q(t)$ is a  parameter defining $\rho$(t) and is not an independent parameter. This type of structure for the expression of the density is also present in class of shear-free solutions (Thompson and Whitrow \cite{tw67}, Bondi \cite{bh69}). Thus, this is a general feature,  irrespective of any assumption. However, the requirement for the density to remain positive throughout the motion yields the constraint $Q(t)<F^{2}(t)/27$.
\par
 Equation (\ref{eq10}) after substituting for $\dot{\lambda}$ from Eq. (\ref{eq20}) leads to
\begin{equation}
p+\rho=-\frac{\dot{\rho}}{\dot{\mu}}[\frac{2}{3}+ \mathcal{G} e^{-\frac{3\mu}{2}}]
\label{eq30}
\end{equation}
The functions $\lambda$  and $\mu$ are determined in conformity with Eqs. (\ref{eq27}) and (\ref{eq20}). Once ${\mu}$ is determined,  ${\nu}$ is given via (\ref{eq18}). The density and the pressure of the perfect fluid distribution are given via Eqs. (\ref{eq28}) and (\ref{eq30}), respectively. The shear of the four-velocity vector using Eqs. (\ref{eq3001}),  (\ref{eq18}) and   (\ref{eq19}) is obtained as
\begin{equation}
\sigma = | \frac{\sqrt{3}}{2}G(t)e^{-\frac{3\mu}{2}} |
\label{eq38}
\end{equation}\par
Let us recapitulate the results obtained so far. There are four functions of time viz., $F$, $G$, $M$ and $Q$; all appearing as  constants of integrations. However, all are not independent of each other. $F(t)$ represents the expansion of the four-velocity of the perfect fluid, whereas $Q(t)$ defines the density $\rho$(t) of the fluid through a linear relation involving $F^{2}$(t). The function $G(t)$ determines shear via Eq. (\ref{eq38}), whereas $M(t)$ occurs in the expression of the mass function $m(r, t)$. Besides, there is functional relationship between $M$ and $\rho$ defined in terms of $G(t)/F(t)$. Hence, we interpret $M$ $(\rho)$ as the generator of the shear of the four-velocity of the fluid. For shear-free motion $M$ will be independent of $\rho$, requiring it to be a constant. Any class of solutions is characterized by the functional relationship $M$=$M(\rho)$, whereas a specific solution will be characterized by the arbitrary functions  $F (t)$ and $Q (t)$. These functions are needed as input for any model one wishes to discuss. In the next Section we present a class of solutions under the assumption that $M$ is a linear function of $\rho$.

\section{Class of Solutions, $M= M_{0} + (2 \pi\mathcal{G}_0) \rho$}\label{sec4}
In order to obtain a solution one has to solve Eqs. (\ref{eq20}) and (\ref{eq27}). However, it seems unlikely that these equations would be solved in its complete generality. One has to make some simplifying assumptions so as to de-couple these equations. Here, we discuss a situation in which we are able to obtain analytical results. This case is defined as
\begin{equation}
M=M_{0}+2\pi\mathcal{G}_{0} \rho,~~~\mathcal{G}(t)=constant=\mathcal{G}_{0}
\label{eq1004}
\end{equation}
This class arises, if one introduces the assumption that $M$ is a linear function of $\rho$.  Equation (\ref{eq500}) then fixes the relationship of $M$ and $\rho$. Equation (\ref{eq20}) may now be integrated as to yield
\begin{equation}
e^{\frac{\lambda}{2}}=B(r)e^{\frac{\mu}{2}}(1+\frac{3}{2} \mathcal{G}_{0}~e^{\frac{-3\mu}{2}})
\label{eq44}
\end{equation}
where $B$($r$) is an arbitrary function of integration.
We employ the  freedom of re-defining  the $r$- coordinate as
\begin{equation}
e^{\frac{\lambda}{2}}dr = e^{\frac{\overline{\lambda}}{2}}~d\overline{r},~~\frac{d\overline{r}}{\overline{r}}=B(r)dr
\end{equation}
Hereafter, unless stated otherwise,  we drop bars over $\lambda$ and  $r$, and also use  prime to denote the  derivative  with respect to $\overline{r}$. This will lead to 
\begin{equation}
 e^{\frac{\lambda}{2}}= \frac{1}{r}~e^{\frac{\mu}{2}}(1+\frac{3}{2} \mathcal{G}_{0}~e^{\frac{-3\mu}{2}})
\label{eq45}
\end{equation} 
 Equation (\ref{eq27}),  using Eqs. (\ref{eq28}) and (\ref{eq1004}),  leads to
\begin{equation}
\frac{e^{-\lambda}\mu'^{2}}{4}=e^{-\mu}+\frac{F^{2}(t)}{9}(1+\frac{3}{2} \mathcal{G}_{0}~e^{\frac{-3\mu}{2}})^{2} - 8~\pi~\frac{\rho}{3}~(1 + \frac{3}{2}~\mathcal{G}_{0}~e^{\frac{-3\mu}{2}}) - 2M_{0}~e^{\frac{-3\mu}{2}}
\label{eq46}
\end{equation}
Here, it is important to note that we have applied a r- coordinate transformation and set $B(r)$=$\frac{1}{r}$. However, Eq. (\ref{eq27}), because of the  combination $e^{-\lambda}~\mu'^{2}$, is unaffected by any such coordinate transformation. 
Substituting the value of $e^{\frac{\lambda}{2}}$ from Eq.(\ref{eq45}), we have
\begin{equation}
\frac{e^{\frac{-\mu}{2}}\frac{\mu'}{2}}{Z[e^{-\mu}+\frac{F^{2}(t)}{9}Z^{2}- 8\pi\frac{\rho}{3}Z-2M_{0}e^{\frac{-3\mu}{2}}]^{\frac{1}{2}}}=\frac{1}{r}
\label{eq48}
\end{equation}
\begin{equation}
Z(\mathcal{G}_{0}, \mu)\equiv(1+\frac{3}{2}\mathcal{G}_{0}e^{\frac{-3\mu}{2}})
\end{equation}
Equation (\ref{eq48}) may be integrated to express that
\begin{equation}
\mbox{some function of}~ [e^{\frac{\mu}{2}},~ F,~ \mathcal{G}_{0}, M_{0},~ \rho]= \ln~r + t
\label{eq49}
\end{equation}
Here, in view of the fact that so far the time coordinate is arbitrary, we have chosen the arbitrary constant of integration as $t$. We may revert this to write as
\begin{equation}
e^{\frac{\mu}{2}}= e^{\frac{\mu}{2}}(F, \mathcal{G}_{0}, M_{0},~\rho, v) ;~\quad    v=  \ln~r + t
\label{eq50}
\end{equation}
The  metric may now be expressed as
\begin{equation}
ds^{2}=\frac{\frac{9\dot{\mu^{2}}}{4}}{(1+\frac{3}{2} \mathcal{G}_{0}~e^{\frac{-3\mu}{2}})^{2}}\frac{dt^{2}}{[F(t)]^{2}}-e^{\mu}[(1+\frac{3}{2} \mathcal{G}_{0}~e^{\frac{-3\mu}{2}})^{2}(\frac{dr}{r})^{2}+d\theta^{2}+(\sin^{2}\theta)~ d\phi^{2}]
\label{eq51}
\end{equation}
The function $\mu$ is given via
\begin{equation}
 \int\frac{U'dr}{[1+\frac{3}{2}\mathcal{G}_{0}U^{3}][U^{2}+\frac{1}{4}F^{2}(t)\mathcal{G}_{0}^{2}U^{6}+\frac{F^{2}}{3} \mathcal{G}_{0}U^{3}-2M(t)U^{3}+3Q(t)]^{\frac{1}{2}}}=-v 
\label{eq52}
\end{equation}
\begin{equation}
U=e^{\frac{-\mu}{2}}
\label{eq2601}
\end{equation}\par
This class of solutions represents the class of shear-free solutions for the case 
\begin{equation}
\mathcal{G}_{0} =0 \Rightarrow       \dot{\lambda} = \dot{\mu}             
\end{equation}
The metric for this subclass of solutions may now be expressed in the isotropic coordinates as
\begin{equation}
ds^{2}= e^{\nu}dt^{2}- e^{\omega}[dr^{2}+r^{2}d\theta^{2}+(r^{2}sin^{2}\theta)~ d\phi^{2}]
\label{eq420}
\end{equation} 
\begin{equation}
e^{\nu}=[3\dot{\omega}/2F(t)]^{2},~~e^{\omega}=e^{\mu}/r^{2}
\label{eq475}
\end{equation}
Let us recall that  spherically symmetric perfect fluid distributions studied employing the metric in isotropic coordinates (\ref{eq420}) and assuming comoving coordinates have shear-free motion. These studies have a long history and are rich in rediscoveries (For details the reader may refer to Stephani $et~al.$ \cite{set03}, Srivastava \cite{sdc92}, Krasinski \cite{ka06}). We will discuss some of these points in Sect.\ref{sec6.1}. For sake of further discussion and reference we present below our results,  Eqs. (\ref{eq26}) and (\ref{eq27}) with $G(t)$=0 and $M($t$)=M_{0}$.
\begin{equation}
r^{3}~e^{\frac{\omega}{2}}[\omega''- \frac{(\omega')^{2}}{2} - \frac{\omega'}{r}] = 6~M_ {0}   
\label{eq600}                           
\end{equation}
\begin{equation}
r^{3}~e^{\frac{\omega}{2}}[\frac{(\omega')^{2}}{4}  + \frac{\omega'}{r}] = -2~M_{0} + 3~r^{3}~e^{\frac{3\omega}{2}}~Q(t)
\label{eq601}                   
 \end{equation}   
Solving these equations and using Eq. (\ref{eq28}) we have
\begin{equation}
e^{-\omega}[\omega'' + \frac{(\omega')^{2}}{4} + \frac{2\omega'}{r}] = 9~Q(t) = \frac{1}{3}F^{2}(t) - 8\pi \rho(t)
\label{eq602}
\end{equation}   
We may express $\omega$  in terms of Weierstrass elliptic  function by recasting Eq. (\ref{eq601}) employing the double transformation of variables as (Glass\cite{genj79}),
\begin{equation}
\Gamma = 1/(r~e^{\frac{\omega}{2}}),~~~ y = 2~ln~ r      
\label{eq604}                       
\end{equation}
We get
 \begin{equation}
(\Gamma_y)^{2} - \frac{1}{4}\Gamma^{2}  +(\frac{M_{0}}{2}) \Gamma^{3}   =  \frac{3}{4}~Q (t)   
\label{eq605}  
\end{equation}
It is pointed out that these equations and solutions, Eqs. (\ref{eq420})- (\ref{eq605}) have been obtained earlier by many investigators, to name a few Thompson and Whitrow \cite{tw67}, Bondi \cite{bh69}, Glass \cite{genj79}.  It is useful to express Eq. (\ref{eq600}) using change of variables as 
\begin{equation}
Y=e^{-\omega/2},~~~u=r^{2}
\label{eq625}
\end{equation}
\begin{equation}
 Y_{uu} =-(3/4)~M_{0}~u^{(-5/2)}~Y^{2}
\label{eq860}
\end{equation}
Here and onwards,  a subscript is also used to denote the derivative w. r. t. the index.
\par               
The earliest investigation of spherically symmetric homogeneous density perfect fluid distribution executing shear-free motion  is due to Kustaanheimo \cite{kp47}. 
Incidently, this work had been out of sight of relativists for a long time and has been brought to notice by Kramer $et~al.$ \cite{skmh80}, and further elaborated by Krasinski \cite{ka06}\footnote{ Krasinski \cite{kal} has identified this class of solutions as Qvist /Kustaanheimo class  of solutions with the remark that  ``{\it The first to consider this case was Qvist (1947); his
presentation was developed and explained by Kustaanheimo (1947).}"  We find that Qvist \cite{qb47} has explored in depth the structure of the field equations for spherically symmetric Vacuum spacetime but has not discussed about homogeneous density fluid. Besides, it  is also true that Kustaanheimo \cite{kp47} in his investigation made use of Qvist's work \cite{qb47}. 
But giving credit to Qvist is unjustified. However, it is remarked that Kustaanheimo \cite{kp47}  made a reference of Qvist's work \cite{qb47} in such a manner that might have been a possible reason for the misinterpretation.}.
 Kustaanheimo \cite{kp47} first considered the fluid with anisotropic pressure and at a later stage specialized to perfect fluids. In order to understand this study properly we analyse in the next Section shear-free fluids with anisotropic pressure.
\section{Anisotropic Fluids Executing Shear-Free Motion}\label{sec5}
We first present Kustaanheimo's \cite{kp47} analysis.
Kustaanheimo \cite{kp47} assumed  (i) the metric in isotropic coordinates (Eq.(\ref{eq420})), and (ii) the energy momentum tensor as given by Eq. (\ref{eq3}),
and obtained the following results\footnote{ It is pointed out that we have changed his notations and conventions to that of ours 
so as to  lead easy correlation to the present work.} 
\begin{equation}
 e^{-\nu}\dot{\omega}^{2}= \frac{4}{3} A(t)
\label{eq56}
\end{equation} 
\begin{equation}
 e^{-\omega}(\omega''+\frac{\omega'^{2}}{4}+\frac{2\omega'}{r})= A(t) - 8\pi\rho(r,t)
\label{eq57}
\end{equation}
\begin{equation}
\frac{1}{2}~p_{1} + p_{2} + \frac{3}{2}~\rho+ \frac{\dot{\rho}}{\dot{\omega}} = 0
\label{eq58}
\end{equation}
Here $A($t$)$ is an arbitrary constant of integration arising from $G^{1}\ _{0}=0$. Thereafter, he introduced the assumption of homogeneity of the density of the fluid. At this stage, he found convenient to separate the analysis depending on whether $\rho(t)= or~\neq A(t)/8\pi$, and obtained the integral of  Eq. (\ref{eq57}) as
\begin{equation}
e^{\frac{\omega}{2}}= \frac{2\zeta}{r~\sqrt{A(t)- 8\pi\rho(t)}},~ \frac{1}{r}= \frac{\zeta'}{\sqrt{\frac{4}{3}\zeta^{4}+\zeta^{2}+C(t)\zeta}};~\quad \rho(t) \neq A(t)/8\pi
\label{eq61}
\end{equation}
\begin{equation}
e^{\omega} =[\alpha(t) + \frac{\beta(t)}{r}]^{4};~\quad \rho~(t) = A(t)/8\pi
\label{eq62}
\end{equation}
where $C$, $\alpha$ and $\beta$ are arbitrary functions of integrations. It is to be noted that $\zeta$ is determined after integrating Eq. (\ref{eq61}) using the method of quadrature. It is pointed out that Eq. (\ref{eq57}) for Vacuum spacetime i.e. with $\rho(r, t)=0$ and its corresponding integrals analogous to Eqs. (\ref{eq61} - \ref{eq62}) had been obtained earlier by Qvist \cite{qb47}. Kustaanheimo \cite{kp47} utilized these results to obtain the required generalization for the case $\rho (r, t)=\rho (t)$, and obtained the solution. Finally, applying the condition of isotropic pressure i.e. $p_{1} = p_{2}$  he has shown that
\begin{equation}
C(t)= K \sqrt{A(t) - 8\pi\rho(t)};~ \alpha(t)\beta(t) = constant = (1/2){m_{0}}
\label{}
\end{equation}
where K and $m_{0}$ are arbitrary constants. It is straight-forward to get the correlation of Kustaaheimo's\cite{kp47} and our parameters as  
\begin{equation}
(i)~~A (t) = (1/3)~ F^{2}(t),~~~(ii)~~A(t) - 8\pi\rho~(t)=9~Q(t), 
\label{eq570}
\end{equation}
\begin{equation}
(iii)~~\zeta = 3\sqrt{Q(t)}/ (2\Gamma), ~(iv)~~K = -M_{0},~~(v)~~ {m_{0}} = M_{0}
\end{equation}\par
Let us note that Kustaanheimo's analysis \cite{kp47} needed to deal separately the cases $\rho (t)=or \neq  A(t)/ 8\pi$ leading to two distinct classes of solutions. But our analysis  does not require any such distinction. This assumes special significance as $\rho$  and A are arbitrary functions of time, and in general may vary continuously. Hence, we take up study of shear-free anisotropic pressure homogeneous fluid distributions, and start with Eq. (\ref{eq57}) with $\rho (r, t)=\rho (t)$. It is simple to identify this equation as Eq. (\ref{eq602}), and to note that  it becomes integrable after it is multiplied by  $r^{3}e^{\frac{3\omega}{2}}( \frac{\omega'}{2} + \frac{1}{r})$. One obtains
\begin{equation}
r^{3}e^{\frac{\omega}{2}}[(\omega')^{2}+4\omega'/r] = 12~r^{3}Q(t)~e^{\frac{3\omega}{2}}+ (32/3)~g(t)
\label{eq603}
\end{equation}
where $g(t)$ is an arbitrary function of integration. Using the change of double variables defined via Eq. (\ref{eq604}), we may express this equation in terms of Weierstrass elliptic function as
 \begin{equation}
12(\Gamma_y)^{2} - 3\Gamma^{2}  - 8g(t) \Gamma^{3}   =  9Q (t)   
\label{eq670}  
\end{equation}
Equations (\ref{eq603}) and  (\ref{eq602}) may be used to obtain
\begin{equation}
Y_{uu}= g(t)~u^{- 5/2}~Y^{2},~~~~~~  Y = e^{-\omega/2},~~~~~~~~ u= r^{2}
\label{eq660}
\end{equation}  
This equation in view of  Eq. (\ref{eq860}) leads to 
\begin{equation}
g ( t ) = constant= - ( 3/ 4 ) M_{0}   \Rightarrow  T^{1}\ _ {1}= T^{2}\ _{2} 
\end{equation}\par 
Thus the solution for spherically symmetric homogeneous density fluid distribution executing shear-free motion is expressible in terms of Weierstrass elliptic function. Let us note that the anisotropy of the fluid is characterized by the arbitrary function $g(t)$; in case it is constant, the pressure of the fluid becomes isotropic. Equations (\ref{eq56}) and (\ref{eq570}), in view of Eq. (\ref{eq3001}), reveal that the expansion of the four-velocity of a homogeneous density fluid distribution is uniform. In other words, it means that if the homogeneity of the density and the shear-free conditions are adopted the fluid not necessarily be a perfect fluid. Hence, specific conditions in this regard may be met out by assigning $g(t)$ suitably. These findings are relevant for cosmological applications.
\section{Discussion and Concluding Remarks}\label{sec6}
\subsection{\bf{Historical Priorities and Credentials}}\label{sec6.1}
Investigations of spherically symmetric perfect fluids considered in an isotropic coordinate system comoving with the matter
lead  immediately to  Eq. (\ref{eq475}).  The other equation which draws the attention is  $T^{1}\ _{1}-T^{2}\ _{2}$=0  (normally referred to as arising because of the isotropy of the pressure ).  McVittie \cite{mgc33} realized  that these two equations are sufficient to determine the metric coefficients and  obtained his famous solution. 
Some years later,  Wyman \cite{wm46} in his investigation related to perfect fluid distributions obeying an equation of state has shown that these two equations may be analyzed to obtain a differential equation for $\omega$ as
\begin{equation}
e^{\frac{\omega}{2}}[\omega''- \frac{(\omega')^{2}}{2} - \frac{\omega'}{r}] =\psi(r)
\label{eq700}
\end{equation}
where $\psi (r)$  is an arbitrary function of integration (refer to Eq. (2.9) of his paper). Here, it is worthwhile to remark  that\\
 
\textbf{1.} Wyman \cite{RefW}  in his paper has expressed:  \textit{``~The author would like to thank Professor Tolman for suggesting this problem, and to say that Eq. (2.9) of the present paper was obtained from him in a private conversation."}  Hence the credit of getting equation (\ref{eq700}) should also be given to Tolman.
\par
\textbf{2.} Narlikar \cite{nvv47}  has also reported to have obtained the integral, Eq.(\ref{eq700}) and has remarked:  \textit{``The integral (1) has escaped the attention of previous investigators."} 
The arbitrary function  of integration $ \beta (r)$ occurring in his equation (1) may be identified as  $( 1/2 )~\psi(r)$. He further reported to have obtained the following differential equation for $\beta(r)$
\begin{equation}
-4\pi~\rho'~e^{3\omega/2}=\beta' + 3~\beta/r
\label{eq699}
\end{equation}
This result when applied to homogeneous density distribution means that 
\begin{equation}
\rho' =0 \Rightarrow ~\psi (r)~ r^{3} = constant   
\label{eq799}      
\end{equation}
,which considered in the light of Eq.(\ref{eq600}) renders the identification of this constant as 6$M_{0}$. Thompson and Whitrow \cite{tw67}  have shown that the pressure isotropy condition requires $\rho'~ e^{3\omega/2}$ to be  a function of $r$  only. Taub \cite{tah68} who re-discovered the result (\ref{eq700}) also obtained the result (\ref{eq799}).  Misra and Srivastava \cite{rd73}  have obtained the generalization of  the relation (\ref{eq699}) applicable for  a charged perfect fluid distribution [refer to their Eq.(4.13); the function $h(r)$ occurring therein may be identified as  $(1/3 )~\beta~r^{3}$].
\par
\textbf{3.} The validity of  Eq.(\ref{eq700}) is based on the requirement that  $T^{1}\ _{1} - T^{2}\ _{2}=0$ and $G^{1}\ _{0}=0$, and hence is equally valid for Vacuum spacetime. This is the reason that this equation also figures in the paper by Qvist \cite{RefQ} wherein $\psi (r)$  is to be identified as  $[f ( r) ]^{1/2}$.
\par
Kustaanheimo and Qvist \cite{kq48} extended McVittie's realization by introducing double change of  variables as given by Eq.(\ref{eq625}).
They showed that the above referred two equations may be employed to obtain a differential equation  for $Y$:
\begin{equation}
Y_{uu} = f (u)  Y^{2}  
\label{eq850}                               
\end{equation}
where $f(u)$ is an arbitrary function of integration. They explored the Lie Symmetry of this equation in order to know potential functional forms $f(u)$ so that the equation is integrated. They arrived at 
\begin{equation}
f (u) = ( a u^{2}+ b  u + c )^{- 5/2}
\label{eq1100}
\end{equation}
where $a$, $b$ and $c$ are constants. For further discussion of this aspect reader may refer to Stephani \cite{s83}, Stephani $et~al.$ \cite{RefS}, Srivastava \cite{scd87}.
Comparison of Eqs. (\ref{eq700}) and (\ref{eq850}) reveals that $f (u)=-\psi (r)/8u$. It is pointed out that the only nonvanishing Weyl tensor component  $\Psi_{2}$  is related to the function $f(u)$  as 
\begin{equation}
\Psi_{2} = (4/3)uf (u) Y^{3}
\label{eq1500}                         
\end{equation}
Stephani $et~al.$ \cite{RefS}, Barnes \cite{ba73} . Accordingly, any shear-free class of solutions may be characterized invariantly on the basis of $f(u)$. For homogeneous density perfect fluid distributions we have 
\begin{equation}
\Psi_{2} = -Y^{3}~M_{0}/r^{3}
\label{eq1505}                         
\end{equation}

\subsection {\bf{Mitra's Analyses}}\label{sec6.2}
 Mitra \cite{ma12} has claimed that a spherically symmetric perfect fluid distribution executing shear-free motion with homogeneous density must of necessity have homogeneous pressure as well. Let us consider his equation (36); function $e^{-h(t)}$  may be identified in our notation as $(4/9) F^{2} (t)$.  
\begin{equation}
e^{\nu}dt^{2} = (\dot{\mu})^{2}~e^{h(t)}dt^{2}
\label{eq635}
\end{equation} 
He applies the gauge freedom of the coordinate system  in Eq. (\ref{eq635}) and introduces a new time coordinate,  $t_{\ast}$ as 
\begin{equation}
(dt_{\ast})^{2}= e^{h(t)}dt^{2}
\label{eq655}
\end{equation} 
Thereafter he states: \textit{``Then in an appropriate new coordinate, we can always set}
\begin{equation}
e^{-h(t)} = 32~\pi~\rho(t)/3~~"
\label{eq675}
\end{equation} 
The logic behind this statement is obscure. However, we take this statement at its face value and understand it to mean that through a suitable $t$-coordinate transformation such a setting could be achieved without any loss of generality. But this is not correct because $(i)e^{- h (t)}$  is proportional to the square of the expansion of the four-velocity of the fluid, and $(ii)\rho(t)$  is $T^{0}\ _{0}$ component of $T^{i}\ _{j}$. Hence, in any $t$-coordinate transformation both of these will behave as scalars. Accordingly, the relation as Eq. (\ref{eq675}) cannot be said to have been achieved without any loss of generality. Further, it is to be noted that the setting (\ref{eq675}) is as requiring a relation connecting the physical variable, $\rho(t)$ to kinematical variable related quantity $e^{- h(t)}$. But these are already governed by the field equations as Eq. (\ref{eq28}), and will require (\ref{eq675}) to lead to the choice $Q(t)=0$. This is the root cause of leading to the particular case, $p=0$.  This shows that the function $h(t)$ cannot be assigned arbitrarily by suitably redefining the time coordinate, as Mitra seems to believe.
\par
In another paper Mitra \cite{ma11} makes the even more surprising claim that a homogeneous and isotropic distribution of perfect fluid, $\rho(r, t)=\rho_{0} (t)$ and $p(r, t)=p_{0} (t)$ must of necessity have zero pressure. He establishes the claim by applying two changes of time variable (refer to his equations ( 30) and (33)]:   
\begin{equation}
e^{\nu}dt^{2} =  e^{-2p_{0}(t)/[\rho_{0}(t)+p_{0}(t)]}~~(dt_{\ast})^{2} = (dt_{\ast~\ast})^{2}
\label{eq775}
\end{equation} 
He then argues  \textit{``But since pressure is scalar, one would again have $p (t_{\ast~\ast})=p (t_{\ast}) = 0$
  …''}. In view of the discussion above $(-p)$, being the spatial component of $T^{i}\ _{j}$, is a scalar for any $t$- coordinate transformation. Hence, in no circumstance any $t$- coordinate transformation, what so ever, can make pressure to be zero. 

Thus we conclude that the findings by Mitra either apply to special situations or grossly unfounded and hence, as such do not warrant attention of concern.
\subsection{\bf{Function} $M(t)$}\label{sec6.3}
The function $M(t)$ is a constant of integration and  may be assigned on the basis of the boundary conditions,  if any. Let us discuss a situation where the solution may be assumed to be regular at the origin of the coordinate system. In this case we may apply the conditions of elementary flatness i.e. as~~~ $r\rightarrow 0,~~~~    e^{\mu/2} \rightarrow  0,~~~  e^{\lambda/2}  \rightarrow   e^{\mu/2}~\mu'/ 2.$\\
Equation (\ref{eq27}) now requires 
\begin{equation}
M(t)= 0\footnote{Thompson and Whitrow \cite{tw67} have also established $\dot{\lambda}=\dot{\mu}$, by obtaining the result $M(t)=0$ employing Eq. (\ref{eq23}) with the requirement $m(0, t)=0$. 
},~~~ G = 0~\Rightarrow~~~~  \dot{\lambda}  = \dot{\mu}  
\label{eq880}
\end{equation}
Consequently, the shear of the four-velocity of the fluid vanishes (Misra and Srivastava \cite{rd72}, \cite{rd73}). If the distribution is also considered to be bounded then Eq. (\ref{eq12}) considered by putting $p=0$ at the comoving boundary $r=r_{b}$ may now be integrated to get $m(r_{b}, t)$ to be a constant. It is well known that the matching of the interior metric to the outside Schwarzschild metric results into $m$ $(r_{b}, t)$ = $M_{schwarzchild}$ (Misner and Sharp \cite{ms64}). Thus for a bounded spherical distribution regular up to the centre $m(0, t)$ and $M(t)$ will vanish, and accordingly, $m(r, t)$ is identified as the mass of the distribution. This would also mean that nonzero $M(t)$ will correspond to the situation where the distribution has a singularity at the centre. However, the solution with $G(t)$ $\neq$ 0 will represent the dynamics of a bounded system with having shear and singularity at the centre.\par  
It is pointed out that the result (\ref{eq880}) requires only the regularity of the solutions to hold at the origin of the coordinate system. This makes the result applicable to cosmology as well, Raychaudhury \cite{rak79}. Since $G$ is a function of $t$, and $G=0$ corresponds to shear-free motion, hence the solutions we discussed [Eqs (\ref{eq50} - \ref{eq2601})] will represent distributions with singularity at the origin until $G=0$. This means that the singularity of the distribution and its
 shearing motion are closely related. Thus the class of solutions we discussed is applicable to the motion of a distribution where shear may evolve with time.
\par
The present investigation is testimony to the viability of the proposition that in order to obtain solutions representing shearing motion and /singularity there is a need to have a fresh look to the solutions obtained earlier in the comoving coordinates, with keeping open the issue of regularity of solutions at the origin of the coordinate system . 
\section{Acknowledgment}
We would like to thank anonymous referees of the paper whose critical and detailed comments resulted into the improvement in the format and contents of the paper. We would like to express our thanks to Prof. A. Krasinski for providing us  Qvist's paper \cite{qb47}, and also for other suggestions.  We would also like to thank Prof. J.V. Narlikar for his help and useful suggestions. We would like to acknowledge the support of the library facilities received from IUCAA, Pune, India.                                                                                                  

\end{document}